\documentstyle[preprint,prl,aps]{revtex} % for preprint form

\newcommand{\Tr}{{\rm Tr}\,}
\newcommand{\half}{{1 \over 2}}
\newcommand{\CD}{{\cal D}}

\newcommand{\CZ}{{\cal Z}}
\newcommand{\CO}{{\cal O}}
\newcommand{\CS}{{\cal S}}

\newcommand{\J}{\bar{J}}
\newcommand{\iN}{{1 \over N}}
\newcommand{\p}{\partial}
\newcommand{\bp}{\bar{\partial}}
\newcommand{\dg}{{\dagger}}
\font\mengen=bbm12
\def\ID{\hbox{\mengen 1}}
\font \barfont = bbm11

\newcommand{\C}{\mbox{\bf\barfont C}}
\begin{document}
%\twocolumn
\draft
\title{\begin{flushright}
{\small\hfill AEI-090\\
\hfill hep-th/9809079}\\
\end{flushright}
Master Partitions for Large $N$ Matrix Field Theories}
\author{Matthias Staudacher \footnote{matthias@aei-potsdam.mpg.de }
\footnote{Supported in part by EU Contract FMRX-CT96-0012} }
\address{
Albert-Einstein-Institut, Max-Planck-Institut f\"{u}r
Gravitationsphysik\\ Schlaatzweg 1\\  D-14473 Potsdam, Germany }
%\date{\today}
\maketitle
\begin{abstract}
We introduce a systematic approach for treating the large $N$ limit of
matrix {\it field} theories.
\end{abstract}
%\pacs{PACS numbers: 11.10.Kk, 11.10.St, 11.15.Pg, 11.15.-q, 12.38.Cy}
\vspace{5.5cm}
%{\small \hspace*{-1.7cm}AEI-xxx\\
%\hspace*{-1.7cm} 
%Supported in part by EU Contract FMRX-CT96-0012
%\begin{multicols}{2}
\newpage
\narrowtext
\section{Introduction}

It has been known for thirty years that quantum field theory simplifies
enormously if the number $N$ of internal field components tends
to infinity. In the case were the $N$ components form
a {\it vector} this leads to exact solutions in any dimension of space-time.
For physical applications, ranging from solid state physics to
gauge theories and quantum gravity, 
a different situation is much more pertinent:
The case of $N^2$ internal components that form a {\it matrix}. 
Here exact solutions have only been produced for very low dimensionalities.
It is one of the outstanding problems of theoretical physics 
to extend large $N$ technology to physically interesting 
dimensions. 

In the present article we will be concerned with matrix ``spin systems'', that
is $D$-dimensional Euclidean lattice field theories whose internal degrees of
freedom are hermitian, complex or unitary $N \times N$ matrices.
The idea is to treat the problem by a three step procedure:

\noindent
(1) Eguchi-Kawai reduction: 
Replace the $N=\infty$ field theory by a one-matrix model coupled to  
appropriate constant external field matrices.

\noindent
(2) Character expansion: Express the partition function of the
one-matrix model of (1) as a sum over polynomial representations 
-- labelled by Young diagrams -- of $U(N)$.

\noindent
(3) Saddle point analysis: Find an effective Young diagram that dominates the
partition sum of (2) in the large $N$ limit.

The insight that step (1) is possible is due to Eguchi and Kawai \cite{ek}.
Intuitively it says that, if a saddle point configuration exists at
$N=\infty$, it should be given by a single translationally invariant matrix
(the so-called master field). In practice the reduction is rather subtle,
and we will be using the twisted EK reduction \cite{twisted}
which results in a one-matrix model in external constant fields encoding
the original (discrete) space-time.

Step (2) is novel in this context and is the main focus of the present work. 
The one-matrix model of (1) still has $N^2$ degrees of freedom, and it is
well known that a saddle point for matrix models can only be found
once the degrees of freedom are reduced as $N^2 \rightarrow N$.
The external fields encoding space-time prevent any naive reduction
to the $N$ eigenvalues of the matrix, which is the route of choice for
simpler models without external fields. But it is possible to replace
the matrix integral by a sum over {\it partitions} corresponding
to a sum over all polynomial representations of $U(N)$. The crucial point is
then that one ends up with a kind of one-dimensional spin model in 
Young diagram space with only $N$ variables: the lenghts of the $N$ rows of 
the diagram. 

Step (3) might appear to be an exotic idea: we claim that the
$N =\infty$ ``master field'' can be described by a ``master partition''.
However, it has already been recently
demonstrated in a series of papers \cite{chars1},\cite{chars2}
that certain infinite sums over partitions are dominated by a saddle point
configuration. This led to the solution of matrix models in 
external fields not treatable by any other method. The present
models are more complicated, but not fundamentally different.

The character expansions we find lead to a very interesting and apparently
novel combinatorial problem in Young pattern space (see section 4).
More insight into this problem will be needed in order to proceed with
the final step (3) of our program, the saddle point analysis. We introduce
what we call ``lattice polynomials'' $\Xi_h$,$\Upsilon_h$
which are polynomials in ${1 \over N}$.
They depend on the Young diagram $h$ and the precise nature 
of the space-time lattice.  

It might be objected that the present approach is futile unless one can
demonstrate that the lattice polynomials $\Xi_h$,$\Upsilon_h$ 
can be explicitly computed or at least bootstraped at $N=\infty$.
But there is one important argument against this pessimistic assessment: 
The lattice polynomials $\Xi_h$,$\Upsilon_h$ only depend on the nature
of the lattice but
{\it not} on the local measure of the minimally coupled (matrix) 
spins of the model\footnote{
Except for the global symmetry of the matrix spins. In this paper we
develop the theory in parallel for the case of ${\rm U}(N)$ global 
symmetry (hermitian matrices) and ${\rm U}(N) \times {\rm U}(N)$ symmetry
(complex matrices). The other classical groups could presumably be treated 
as well, but it is well known that they do not lead to different large $N$
limits.}.
Therefore, solving interacting field theory in our language is of the same
degree of complexity as solving the free field case. 

Finally we should mention that our program is very general since
it applies in principle to {\it any} large $N$ matrix spin system. 
It would be interesting to extend the method to matrix field theories with a 
{\it gauge symmetry}
such as Yang-Mills theory. Indeed the EK reduction was initially
designed for lattice gauge theory \cite{ek}. Recently it was demonstrated
by Monte Carlo methods that even the path integral of
continuum gauge theory may be 
EK reduced to a {\it convergent} ordinary multiple matrix integral
\cite{ks}. A rigorous mathematical proof, as well as an investigation on
whether  
the reduced model reinduces the field theory as $N \rightarrow \infty$,
are still lacking. At any rate, reducing a $D$-dimensional gauge theory, 
one so far ends up with a nonlinearly coupled
$D$-matrix model, which is not yet tractable by the present machinery
unless it is understood how to perform a further reduction 
$D N^2\rightarrow N^2$.

\section{Reduced Matrix Spin Systems}

Consider a spin model on a periodic lattice.
In order to be specific we will scetch the method for
a two-dimensional lattice, but higher (or lower)
dimensions can be treated as well.
We will not dwell on details since they are well explained elsewhere.
The variables are $N\times N$ hermitian matrices $M(x)$
defined on the lattice sites $x$
\begin{displaymath}
\CZ_{\rm H}= \int \prod_x \CD M(x)~e^{ -\CS_{\rm H}},
\end{displaymath}
\begin{equation}
\CS_{\rm H}= N \Tr \sum_x \bigg[ \half M(x)^2 + V\big(M(x)\big) - 
 {\beta \over 2} \sum_{\mu=1,2}
[M(x) M(x+\hat{\mu}) + M(x) M(x-\hat{\mu})]  \bigg],
\label{mlattice}
\end{equation}  
where $\hat{\mu}$ denotes the unit vector in the $\mu$-direction.
It is equally natural to consider general complex matrices 
$\Phi(x) \in {\rm GL}(N,\C)$, in which case
\begin{displaymath}
\CZ_{\rm GL}= \int \prod_x \CD \Phi(x)~e^{ -\CS_{\rm GL}},
\end{displaymath}
\begin{equation}
\CS_{\rm GL}= N \Tr \sum_x \bigg[
\Phi(x) \Phi^\dg(x) + V\big(\Phi(x) \Phi^\dg(x)\big)
- \beta \sum_{\mu=1,2} [\Phi(x) \Phi^\dg(x+\hat{\mu}) + 
\Phi(x) \Phi^\dg(x-\hat{\mu})] \bigg].
\label{philattice}
\end{equation}
If $V=0$ in eqs.(\ref{mlattice}),(\ref{philattice}) the model is free. 
The integration measures in 
eqs.(\ref{mlattice}),(\ref{philattice}) are the flat measures for 
hermitian and complex matrices: 
\begin{equation}
\CD M=\prod_{i=1}^{N} {dM_{i i} \over \sqrt{2 \pi N^{-1}}} 
\prod_{i<j}^N {d{\rm Re}M_{i j} d{\rm Im}M_{i j} \over \pi N^{-1}},
\;\;\;\;\;\;\;\;
\CD \Phi=\prod_{i,j=1}^{N}
{d{\rm Re}\Phi_{i j} d{\rm Im}\Phi_{i j} \over \pi N^{-1}}. 
\label{measures}
\end{equation}
A third, very important type of spin model is the so-called 
{\it chiral field}, which looks like the free complex model 
eq.(\ref{philattice})
\begin{displaymath}
\CZ_{\rm U}= \int \prod_x \CD U(x)~e^{ -\CS_{\rm U}}, 
\end{displaymath}
\begin{equation}
\CS_{\rm U}= -\beta N \Tr \bigg[\sum_x 
\sum_{\mu=1,2}
[U(x) U^\dg(x+\hat{\mu}) + 
U(x) U^\dg(x-\hat{\mu})] \bigg].
\label{ulattice}
\end{equation}  
but the matrices $U(x) \in {\rm U}(N)$ are unitary. In this case 
the measure $\CD U(x)$ is the Haar measure on the group. 
The model is therefore not free.

The Eguchi-Kawai reduction \cite{ek},\cite{twisted} 
states that the above lattice models
can be replaced at $N=\infty$ by, respectively, the
following one-matrix models coupled to constant external field matrices
$P$ and $Q$:
\begin{equation}
Z_{\rm H}= \int \CD M \exp 
N \Tr \bigg[
-\half M^2 - V\big(M\big) +\beta \Big( M P M P^\dg + M Q M Q^\dg \Big)  \bigg],
\label{mtwist}
\end{equation}  
\begin{eqnarray}
\lefteqn{Z_{\rm GL}= \int \CD \Phi~\exp N \Tr \Big[ 
-\Phi \Phi^\dg  - V\big(\Phi \Phi^\dg\big) \Big] \times } \cr 
& \quad \quad \quad \quad \quad &
\times \exp \beta N \Tr \Big( \Phi P \Phi^\dg P^\dg 
+ \Phi P^\dg \Phi^\dg P + \Phi Q \Phi^\dg Q^\dg + \Phi Q^\dg \Phi^\dg Q \Big),
\label{phitwist}
\end{eqnarray}
\begin{equation}
Z_{\rm U}= \int \CD U \exp~\beta N \Tr \Big( 
U P U^\dg P^\dg + U P^\dg U^\dg P + U Q U^\dg Q^\dg + U Q^\dg U^\dg Q \Big).
\label{utwist}
\end{equation} 
Here $P=P_N$ and $Q=Q_N$ are the famous $N \times N$
unitary ``shift and clock'' matrices
\begin{equation}
P_N=\left( \begin{array}{cccccc}
0 & 1 &   &  &  &  \\
  & 0 & 1 &  &  &   \\
  &   & \ddots & \ddots &  &  \\
  &   &        &        & 0  & 1\\
1 &  & & &  & 0
\end{array} \right),
\quad Q_N=\left( \begin{array}{cccccc}
1 &  &   &  &  &  \\
  & \omega_N &   &  &  &   \\
  &   & \ddots &  &  &  \\
  &   &        &        & \omega_N^{N-2}  & \\
  &  & & &  & \omega_N^{N-1}
\end{array} \right),
\label{pq}
\end{equation}
where $\omega_N=\exp {2 \pi i \over N}$ and $P_N Q_N=\omega_N Q_N P_N$.
To be more precise, the free energies as well as appropriate
correlation functions (see \cite{twisted}) are identical to leading order in  
$\iN$ in the lattice field theory and the corresponding one-matrix model.
The thermodynamic limit, that is a lattice of infinite extent, is
approached when $N \rightarrow \infty$. 
We see that the structure of the lattice has been ``hidden'' in 
index space!
It is natural to generalize
the situation to a toroidal $K \times L$ lattice: 
\begin{equation}
P=P_K \otimes \ID_{N \over K}, \quad \quad Q=Q_L \otimes \ID_{N \over L},
\label{finite}
\end{equation}
where $N$ is chosen to be divisible by $K$ and $L$. This allows to take
the thermodynamic limit and the large $N$ limit independently.
If we put $L=1$ (we can then equivalently omit $Q$ altogether) the target
space becomes a closed one-dimensional chain.

We suspect that matrix models on arbitrary discrete target spaces
can be EK reduced by appropriate external matrices, but this has not
been worked out yet.

\section{Character Expansions}

Now we turn to step (2) and rewrite the reduced hermitian, complex and
unitary matrix integrals eqs.(\ref{mtwist}),(\ref{phitwist}),(\ref{utwist})
as sums over representations of U$(N)$. To this end introduce the
following source integrals:
\begin{equation}
Z_{\rm H}[J]=\int \CD M~\exp~N \Tr \Big[ -\half M^2 -V(M) + J M \Big],
\label{msource}
\end{equation}
\begin{equation}
Z_{\rm GL}[J \J]=\int \CD \Phi~\exp~N 
\Tr \Big[ -\Phi \Phi^\dg -V(\Phi \Phi^\dg) +
J \Phi + \Phi^\dg \J \Big],
\label{phisource}
\end{equation}
\begin{equation}
Z_{\rm U}[J \J]=\int \CD U~\exp~N \Tr \Big[J U + U^\dg \J \Big].
\label{usource}
\end{equation}
The two different ways of introducing a source are due to
the U$(N)$ symmetry of hermitian matrices on the one hand
and the ${\rm U}(N) \times {\rm U}(N)$ symmetry of complex (and complex
unitary) matrices on the other.
The reduced models are easily obtained from the source integrals by
applying an operator:
\begin{equation}
Z_{\rm H}=\exp {\beta \over N} \Tr \Big(\p P \p P^\dg + \p Q \p Q^\dg \Big) 
\cdot Z_{\rm H}[J]~\Big|_{J=0},
\label{mop}
\end{equation}
\begin{equation}
Z_{\rm GL,U}=\exp {\beta \over N} \Tr \Big(\p P \bp P^\dg + \p P^\dg \bp P +
\p Q \bp Q^\dg + \p Q^\dg \bp Q \Big) \cdot Z_{\rm GL,U}[J \J]~\Big|_{J=\J=0}.
\label{phiuop}
\end{equation}
Here $\p$,$\bp$ denote $N \times N$ matrix differential operators
whose matrix elements are $\p_{j i}={\p \over \p J_{i j}}$ and
$\bp_{j i}={\p \over \p \J_{i j}}$. It is clear that the source integrals
are {\it class} functions of, respectively, $J$ and $J \J$. Therefore
they can be expressed as character expansions, 
with known (see \cite{chars1},\cite{chars2},\cite{bars}) expansion 
coefficients.
If $V=0$, they read
for the hermitian and complex source integrals, respectively,
\begin{equation}
Z_{\rm H}[J]=\exp~ \half N \Tr J^2 = \sum_h \chi_h(A_2) \chi_h(J),
\label{mj}
\end{equation}
\begin{equation}
Z_{\rm GL}[J \J]=\exp~N \Tr J \J=\sum_h \chi_h(A_1) \chi_h(J \J),
\label{phij}
\end{equation}\
while for the unitary source integral one has \cite{bars}
\begin{equation}
Z_{\rm U}[J \J]=\sum_h {\chi_h(A_1) \chi_h(A_1) \over \chi_h(\ID)} 
\chi_h(J \J).
\label{uj}
\end{equation}
Here the sum runs over all partitions $h$ labeled by the shifted
weigths $h_i=N-i+m_i$, where $m_i \geq 0 $, 
$i=1, \ldots, N$, is the number of boxes in the $i$'th row of the Young
pattern associated to $h$. 
$\chi_h(J)$ is the Schur function, dependent on $J$, on the
diagram $h$. 
It is identical to the Weyl character of the matrix $J$ corresponding
to the representation labeled by $h$.
$A_1$ and $A_2$ are defined through 
$\Tr A_1^k=N(\delta_{k,0} + \delta_{k,1})$ and 
$\Tr A_2^k=N(\delta_{k,0}+\delta_{k,2})$, and $\chi_h(\ID)$ is the 
dimension of the representation. 
For more details on the notation,
and for explicit formulas for the characters 
$\chi_h(A_1)$, $\chi_h(A_2)$ and $\chi_h(\ID)$ see 
\cite{chars1},\cite{chars2}.
For a non-zero potential $V$, the hermitian and complex
character expansions become a bit more complicated, but are still available:
\begin{equation}
Z_{\rm H}[J]= \sum_h \Theta_h \chi_h(J),
\label{mjfull}
\end{equation}
\begin{equation}
Z_{\rm GL}[J \J]=\sum_h \Omega_h \chi_h(J \J),
\label{mphifull}
\end{equation}
where $\Theta_h$ is given by 
\begin{equation}
\Theta_h={\chi_h(A_1) \over \chi_h(\ID)}~
\int \CD M~\exp~N \Tr \Big[ -\half M^2 -V(M)\Big]~\chi_h(M),
\label{hermitian}
\end{equation}
and $\Omega_h$ by
\begin{equation}
\Omega_h=\Bigg({\chi_h(A_1) \over \chi_h(\ID)}\Bigg)^2~
\int \CD \Phi~\exp~N \Tr \Big[ -\Phi \Phi^\dg -V(\Phi \Phi^\dg) \Big]
~\chi_h(\Phi \Phi^\dg).
\label{complex}
\end{equation}
The integrals appearing in eqs.(\ref{hermitian}),(\ref{complex}) are
ordinary one-matrix integrals which may be computed rather explicitly
as $N \times N$ determinants.
Their analysis in the $N \rightarrow \infty$ 
limit proceeds by employing  standard techniques, supplemented by the
methods of \cite{chars1}.

Now we apply the operators in eqs.(\ref{mop}),(\ref{phiuop}) in order to
generate the space-time lattice; this results in character expansions
for the reduced matrix field theories. In the hermitian case one has
(here $|h|=\sum_i m_i=$number of boxes in the Young diagram)
\begin{equation}
Z_{\rm H}=\sum_h \chi_h(A_2)~\Xi_h~\beta^{{|h| \over 2}}
\;\;\;\;\;\;{\rm for}\;\;\;\;\;\;V=0,
\label{mexp}
\end{equation}
\begin{equation}
Z_{\rm H}=\sum_h \Theta_h~\Xi_h~\beta^{{|h| \over 2}}
\;\;\;\;\;\;{\rm for}\;\;\;\;\;\;V \neq 0,
\label{mfull}
\end{equation}
with
\begin{equation}
\Xi_h=\exp\iN \Tr \Big(\p P \p P^\dg + \p Q \p Q^\dg \Big) 
\cdot \chi_h(J)~\Big|_{J=0}.
\label{Xi}
\end{equation}
The free complex, interacting complex, and the unitary case become
\begin{equation}
Z_{\rm GL}=\sum_h \chi_h(A_1)~\Upsilon_h~\beta^{|h|} 
\;\;\;\;\;\;{\rm for}\;\;\;\;\;\;V=0,
\label{phiexp}
\end{equation}
\begin{equation}
Z_{\rm GL}=\sum_h \Omega_h~\Upsilon_h~\beta^{|h|} 
\;\;\;\;\;\;{\rm for}\;\;\;\;\;\;V \neq 0,
\label{phifull}
\end{equation}
\begin{equation}
Z_{\rm U}=
\sum_h {\chi_h(A_1) \chi_h(A_1) \over \chi_h(\ID)}~\Upsilon_h~\beta^{|h|},
\label{uexp}
\end{equation} 
with
\begin{equation}
\Upsilon_h=\exp\iN \Tr \Big(\p P \bp P^\dg + \p P^\dg \bp P +
\p Q \bp Q^\dg + \p Q^\dg \bp Q \Big) \cdot \chi_h(J \J)~\Big|_{J=\J=0}.
\label{Upsilon}
\end{equation}
The character expansions 
eqs.(\ref{mexp}),(\ref{mfull}),(\ref{phiexp}),(\ref{phifull}),(\ref{uexp}) 
are at the heart of our proposal. It is seen
that they neatly {\it separate} the nature of the local spin weight 
($\chi_h(A_2)$,$\Theta_h$,$\chi_h(A_1)$,$\Omega_h$,
$(\chi_h(A_1))^2 (\chi_h(\ID))^{-1}$)
and the nature of the embedding space ($\Xi_h$,$\Upsilon_h$). 
As a striking example, note that from the
point of view of our character expansion method the
difference between the free Gaussian model on a toroidal lattice 
eq.(\ref{phiexp}) and the non-trivial chiral model eq.(\ref{uexp}) 
is a simple, explicitly known factor 
\begin{displaymath}
{\chi_h(A_1) \over \chi_h(\ID)}= N^{|h|} \prod_{i=1}^N {(N-i)! \over h_i!}. 
\end{displaymath}
The character expansions involve sums over $N$ variables only and we can
write down a saddle point equation for the effective density of
the master partition. In order to complete the program, we need a 
second bootstrap equation for the novel quantities $\Xi_h$ and $\Upsilon_h$,
which contain the connectivity information of the lattice.

\section{Lattice Polynomials}

Inspection of the quantities $\Xi_h$ and $\Upsilon_h$ in 
eqs.(\ref{Xi}),(\ref{Upsilon}) shows that they are polynomials in the
variable ${1 \over N}$ of degree not higher than, respectively,
$\half |h| -1$ and $|h|-2$. They are zero if the number $|h|$ of boxes in the 
Young pattern is odd. 
Conjugating the diagram gives the same polynomial except for 
the replacement $\iN \rightarrow -\iN$. 
The first few can be computed by brute force calculation directly from
the definitions eqs.(\ref{Xi}),(\ref{Upsilon}), see Table. 

\begin{center}
Table : The first few $D=2$ lattice polynomials \\
\vspace{0.5cm}
\begin{tabular}{|l||l|l||} \hline
$h$ & $ \Xi_h $     &  $ \Upsilon_h $       \\ \hline \hline
$ 2 $ & $ 2 $  &  $ 2 $    \\
$ 1^2 $ & $ 2 $  &  $ 2 $        \\ \hline
$4$ & $ 3+ 12 \iN $  &  $ 3+24 \iN +54 {1 \over N^2} $            \\
$3 1 $ & $ 5+ 4 \iN$  &  $5+8 \iN+18 {1 \over N^2} $           \\
$2^2$ & $ 6 $  &  $ 6  $          \\ 
$2 1^2$ & $ 5- 4 \iN $  &  $5-8 \iN+18 {1 \over N^2} $          \\
$1^4$ & $ 3- 12 \iN $  &  $3-24 \iN +54 {1 \over N^2} $          \\
\hline
\end{tabular} 
\end{center}

Here we used
$\Tr (P^k Q^l)=N \delta_{k,0} \delta_{l,0}$, which is true as long as
$|k|<N$, $|l|<N$. We also replaced 
$\omega_N \rightarrow 1$, $\omega_N^* \rightarrow 1$ (remember 
$\omega_N=\exp {2 \pi i \over N}$): in other words, we assumed
$P$ and $Q$ to commute at large $N$. Both assumptions are innocent 
at least in the strong coupling (small $\beta$) phase.
If the model  possesses a weak coupling phase (like e.g.~the chiral 
field eq.(\ref{utwist})), these assumptions
may have to be reconsidered, if we want the character expansion to describe
this second phase as well. This is because in the present approach
we expect large $N$ phase transitions to correspond to the
situation where the number of rows of the master partition is of $\CO(N)$
(``touching transition'').
Note that we {\it cannot} drop
the other terms of $\CO(\iN)$ in $\Xi_h$,$\Upsilon_h$ since the character
expansions are for the partition function and not for the free energy. 

The direct calculation of the lattice polynomials quickly gets very tedious.
The combinatorics involved seems to be of a novel type. 
While we have not yet found 
an efficient calculational scheme or recursive method, let us give some
interesting representations for $\Xi_h$ and $\Upsilon_h$ that may prove
useful later. Introduce the following Gaussian measure on the space of
$M \times N$ ($M \leq N$) complex matrices $\Lambda$:
\begin{equation}
[\CD \Lambda]=\prod_{i=1}^M \prod_{j=1}^{N} \Bigg(
{d{\rm Re}\Lambda_{i j} d{\rm Im}\Lambda_{i j} \over \pi N^{-1}} \Bigg)
~\exp N \Tr \Big[ -\Lambda \Lambda^\dg \Big].
\label{rechteck}
\end{equation}
This measure is invariant under ${\rm U}(M) \times {\rm U}(N)$.
It is then fairly easy to prove ({\it cf} \cite{chars2})
the following representation for the
character of the source:
\begin{equation}
\chi_h(J)=\int \CD U \chi_h(U^\dg)~\int [\CD \Lambda]
~\exp N \Tr U \Lambda J \Lambda^\dg,
\label{source}
\end{equation}
where $U\in{\rm U}(M)$ is unitary and $\CD U$ is the Haar measure on U$(M)$.
This formula is valid for diagrams $h$ with at most $M$ rows.
Therefore $\Xi_h$ becomes, {\it cf} eq.(\ref{Xi})
\begin{equation}
\Xi_h=\int \CD U \chi_h(U^\dg)~\int [\CD \Lambda]~
\exp N \Tr \Big( \Lambda^\dg U \Lambda P \Lambda^\dg U \Lambda P^\dg +
\Lambda^\dg U \Lambda Q \Lambda^\dg U \Lambda Q^\dg \Big).
\label{cute}
\end{equation}
After a Hubbard-Stratanovich transformation decoupling the quartic terms
by Gaussian $M \times M$ complex matrices $S$ and $T$ (with measure as
in eq.(\ref{rechteck}) with $N \rightarrow M$), and integration over
$\Lambda$, we obtain the representation
\begin{eqnarray}
\lefteqn{\Xi_h=\int \CD U \chi_h(U^\dg)~\int [\CD S] [\CD T] \times} \cr
& \quad \quad \quad  &
\times \exp \bigg[ \Tr_{M \otimes N} \sum_{k=1}^{\infty} {1 \over k} 
\Big(S U \otimes P
+S^\dg U \otimes P^\dg + T U \otimes Q + T^\dg U \otimes Q^\dg \Big)^k \bigg].
\label{walks}
\end{eqnarray}
The combinatorial interpretation of the exponential in eq.(\ref{walks})
is the following: we have a generating function for a non-commutative 
random walk on a two-dimensional lattice with variable $U$.
The representation is useful for getting some exact results on the 
$\Xi_h$, but we have not yet been able to compute the integral 
eq.(\ref{walks}) exactly except for $M=1$ (characters with just one row). 
E.g.~we can find a generating function (with $z_i$ being the eigenvalues
of $U$) for the large $N$ limit of $\Xi_h$
\begin{equation}
\prod_{i,j}^M {1 \over (1- z_i z_j)^2}=\sum_h \Xi_h^{N=\infty} \chi_h(z)
\label{constant}
\end{equation}
giving the constant terms of the lattice polynomials. This is however not
sufficient for the large $N$ limit of the field theory, as already 
mentioned. A curious feature of eq.(\ref{walks}) is that we can
take $N \rightarrow \infty$ while keeping $M$ in the range
$1 \ll M \ll N$. That is,
it should be possible to find a saddle point for the situation where
the row lenghts are large compared to the number of rows, corresponding
to  the extreme strong coupling limit. Furthermore, it should be
investigated whether the $M \times M$
matrices can be taken to commute as $N \rightarrow \infty$.

Similar, if slightly more complicated representations are possible
for $\Upsilon_h$; here the starting point is the expression
\begin{equation}
\chi_h(J \J)=\int \CD U \chi_h(U^\dg)~\int [\CD \Lambda_1] [\CD \Lambda_2]
~\exp N \Tr \Big( U^{\half} \Lambda_1 J \Lambda_2^\dg 
+ \Lambda_2 \J \Lambda_1^\dg U^{\half} \Big),
\label{doublesource}
\end{equation}
which means the lattice polynomials become
\begin{eqnarray}
\lefteqn{\Upsilon_h=\int \CD U \chi_h(U^\dg)~\int [\CD \Lambda_1] 
[\CD \Lambda_2] \times} \cr
& \quad \quad \quad  &
\times \exp N \Tr \Big( \Lambda_2^\dg U^\half \Lambda_1 P \Lambda_1^\dg 
U^\half  \Lambda_2 P^\dg +
\Lambda_2^\dg U^\half \Lambda_1 P^\dg \Lambda_1^\dg 
U^\half  \Lambda_2 P \Big) \times \cr
& \quad \quad \quad  &
\times \exp N \Tr \Big(
\Lambda_2^\dg U^\half \Lambda_1 Q \Lambda_1^\dg U^\half \Lambda_2 Q^\dg +
\Lambda_2^\dg U^\half \Lambda_1 Q^\dg \Lambda_1^\dg U^\half \Lambda_2 Q \Big),
\label{cutezwei}
\end{eqnarray}
and the non-commutative random walk representation is
\begin{eqnarray}
\lefteqn{\Upsilon_h=\int \CD U \chi_h(U^\dg)~\int 
[\CD S] [\CD \bar{S}] [\CD T] [\CD \bar{T}] \times} \cr
& \quad \quad \quad  &
\times \exp \bigg[ \Tr_{M \otimes N} \sum_{k=1}^{\infty} {1 \over k} 
\Big(S U^{\half} \otimes P
+\bar{S} U^{\half} \otimes P^\dg + T U^{\half} \otimes Q 
+ \bar{T} U^{\half} \otimes Q^\dg \Big)^k \bigg] \times \cr
& \quad \quad \quad  &
\times \exp \bigg[ \Tr_{M \otimes N} \sum_{k=1}^{\infty} {1 \over k} 
\Big(\bar{S}^\dg U^{\half} \otimes P
+S^\dg U^{\half} \otimes P^\dg + \bar{T}^\dg U^{\half} \otimes Q 
+ T^\dg U^{\half} \otimes Q^\dg \Big)^k \bigg],
\label{walks2}
\end{eqnarray}
from which we find that $\Upsilon_h^{N=\infty}=\Xi_h^{N=\infty}$, 
{\it cf} eq.(\ref{constant}), but ${1 \over N}$ corrections are
different (see Table above). Again, for arbitrary one-row
representations ($M=1$) it is possible to obtain $\Upsilon_h$ rather
explicitly.

Another potentially useful representation\footnote{We thank D.~Verma
for pointing this out to us.} of the lattice polynomials
is given by the following {\it dual} equations: eq.(\ref{Xi}) 
becomes
\begin{equation}
\Xi_h=\chi_h(\p) \cdot
\exp\iN \Tr \Big(J P J P^\dg 
+ J Q J Q^\dg \Big)~\Big|_{J=0}, 
\label{Xidual}
\end{equation}
and eq.(\ref{Upsilon}) is dual to
\begin{equation}
\Upsilon_h=\chi_h(\p \bp) \cdot
\exp\iN \Tr \Big(J P \J P^\dg + J P^\dg \J P +
J Q \J Q^\dg + J Q^\dg \J Q \Big)~\Big|_{J=\J=0}.
\label{Upsilondual}
\end{equation}

We could go on and discuss correlation functions which are naturally
included into the present formalism. In particular, it is straightforward
to give expressions for their character expansions in terms of modified
lattice polynomials, and it remains true that the combinatorics
is independent on whether the reduced field theory is free or interacting. 
This is however beyond the scope of the present article.

While it is unclear whether the $D \geq 2$ lattice polynomials can be 
computed exactly for a general partition, it should be stressed once more 
that this is unnecessary;
all we need is an indirect method in order to extract the large $N$ behavior.

\section{Conclusions}

This solution to the problem of the large $N$ limit
of (non-gauge) matrix field theories is not yet complete
since the structure of the lattice 
polynomials we introduced still needs to be further analyzed
in order to be able to write the full set of saddle point equations.
However we feel that we are definitely closing in on the large $N$
problem, and that we have brought it into the simplest form to date.
The proposed approach is concrete, systematic 
and rather general: we demonstrated that the reduction from $N^2$
to $N$ variables is possible once one changes
variables from matrices to partitions. In this language the
{\it master field} becomes a {\it master partition}.
Presumably one should first (re)derive in the current framework
the exact solutions for some lower dimensional target spaces before 
dealing with the two (and higher) dimensional field theories.

\acknowledgements
We thank Dayanand Verma and Brian G.~Wybourne for interesting and
useful discussions concerning the combinatorial aspects of this project.
This work was supported in part by the EU under Contract FMRX-CT96-0012.

\end{document}